\documentclass[aps,pra,preprint,superscriptaddress]{revtex4-1}

\usepackage{amsmath}
\usepackage{amssymb}
\usepackage{amsthm}
\usepackage{mathrsfs}
\usepackage{graphicx}
\usepackage{color}
\usepackage{enumitem}
\usepackage{caption}
\usepackage{subfigure}

\begin{document}


\title{Quantum algorithm and quantum circuit for A-Optimal Projection: dimensionality reduction}
\author{Bojia Duan}
\email[]{deja@nuaa.edu.cn}
\affiliation{College of Computer Science and Technology, Nanjing University of Aeronautics and Astronautics, No.29 Jiangjun Avenue, 211106 Nanjing, China.}
\affiliation{Institute for Quantum Computing, University of Waterloo, 200 University Ave W, Waterloo, ON N2L 3G1, Canada.}
\author{Jiabin Yuan}
\email[]{jbyuan@nuaa.edu.cn}
\author{Juan Xu}
\author{Dan Li}
\affiliation{College of Computer Science and Technology, Nanjing University of Aeronautics and Astronautics, No.29 Jiangjun Avenue, 211106 Nanjing, China.}

\date{\today}

\begin{abstract}

Learning low dimensional representation is a crucial issue for many machine learning tasks such as pattern recognition and image retrieval. In this article, we present a quantum algorithm and a quantum circuit to efficiently perform A-Optimal Projection for dimensionality reduction. Compared with the best-know classical algorithms, the quantum A-Optimal Projection (QAOP) algorithm shows an exponential speedup in both the original feature space dimension $n$ and the reduced feature space dimension $k$. We show that the space and time complexity of the QAOP circuit are $O\left[ {{{\log }_2}\left( {nk} /{\epsilon} \right)} \right]$  and $O[ {\log_2(nk)}  {poly}\left({{\log }_2}\epsilon^{-1} \right)]$ respectively, with fidelity at least $1-\epsilon$.
Firstly, a reformation of the original QAOP algorithm is proposed to help omit the quantum-classical interactions during the QAOP algorithm. Then the quantum algorithm and quantum circuit with performance guarantees are proposed. Specifically, the quantum circuit modules for preparing the initial quantum state and implementing the controlled rotation can be also used for other quantum machine learning algorithms.

\end{abstract}


\maketitle


\section{Introduction}
\label{sec1:level1}

     Learning low dimensional image representations has gained significant importance in many image processing tasks such as recognition and retrieval \cite{CJL17, WNH17, SLH17}. A range of applications of this problem can be seen in the field of medical imaging such as liver cirrhosis, lung cancer classification and breast cancer diagnosis \cite{SLH17}. Another typical example is face recognition which is typically used in security systems or as a commercial identification and marketing tool \cite{JHJ18}. 
     Recent studies have shown that images are possibly sampled from a low dimensional manifold, however, the visual features, such as color, texture and shape, which are usually extracted for the image representation, are usually of very high dimensionality \cite{HZZ16}. Therefore, a range of techniques have been developed for dimensionality reduction. For instance, principal component analysis (PCA) is guaranteed in terms of the linearly embedded manifold \cite{Bishop06}. Moreover, Isomap, Locally Linear Embedding, and Laplacian Eigenmap are proposed for nonlinear embedded manifold \cite{TSL00, RS00, BN01}.
     
     Different from all the aforementioned techniques which are not directly related to the regression task, X. He proposed a novel dimensionality reduction algorithm named A-Optimal Projection (AOP) which performs better regression performance in the reduced space \cite{HZZ16}. This approach can be performed under either unsupervised or supervised mode, which is a more widely used algorithm compared to the unsupervised algorithm PCA or the supervised algorithm linear discriminant analysis (LDA) \cite{Bishop06}. Moreover, different from most dimensionality reduction algorithms which are applied as pre-processing of the data, AOP can directly improve the performance of a regression model in the reduced space, therefore, the learned regression model can be as stable as possible. 
     
     Time complexity is a significant drawback in classical machine learning algorithms. A range of quantum algorithms have achieved exponentially speed up in machine learning compared with the classical ones \cite{AAD15, Wittek14}. 
     In particular, quantum algorithms for solving the problem of pattern classification and image classification problems were proposed, covering an important area of machine learning \cite{SSP16, Duan17}.
     Recently, a quantum generative algorithm which is more capable of representing probability distributions was proposed, generating an intriguing link among quantum many-body physics, quantum computational complexity theory and the machine learning frontier \cite{GZD18}.
     The relationship between feature maps, kernel methods in machine learning and quantum computing was also investigated, and the idea of embedding data into a quantum Hilbert space opens up a promising avenue to quantum machine learning \cite{SK19}. 
     Moreover, small quantum computers, larger special purpose quantum simulators, annealers, etc., exhibit promising applications in machine learning, and the perspectives on the work of these hardware have also been discussed \cite{BWP17}.
     Quantum machine learning has also been combined with the information security.  It was designed to protect private data during performing quantum machine learning, which has potential applications in the big data era \cite{SZ17}.
     
     In the application field of quantum dimensionality reduction, the quantum algorithm for PCA has been proposed for unsupervised mode \cite{LMR14}, and the quantum algorithm for LDA has been proposed for supervised mode and classification \cite{CD16}. In this paper, we focus on the new dimensionality reduction algorithm AOP which can be used both on unsupervised and supervised model, and propose a quantum algorithm for AOP, which achieves exponentially speedup compared with the classical polylogarithmic in both $n$, the original feature space dimension,  and $k$, the reduced feature space dimension.     
     
     Our work has two major contributions. 
     First, we present a quantum algorithm for solving the learning process of the AOP algorithm. A reformulation of the original classical AOP algorithm is introduced here which helps the QAOP algorithm be implemented more efficiently. The quantum algorithm is made of iterations, where each iteration mainly consists of phase estimation and a controlled rotation. The reformulated AOP and the partial trace technology can help omit the quantum-classical interactions during the quantum algorithm.
     Second, we design a detailed quantum circuit for the proposed QAOP algorithm which makes it possible to execute the QAOP algorithm on a universal quantum computer. The circuit for preparing the initial state is presented and the detailed circuit for the controlled rotation is designed. The space and time analysis of the quantum circuit also shows an exponential speedup in the size of the feature space than the classical counterparts.
     
     This paper is arranged as follows: We give a brief overview of the classical AOP algorithm in Sec.  \ref{sec2:level1}. In Sec. \ref{sec3:level1}, the quantum algorithm for AOP algorithm which is used in dimension reduction is presented. In Sec. \ref{sec4:level1}, the overview and detailed quantum circuits for solving QAOP algorithm are designed. Finally we show the conclusions in Sec. \ref{sec5:level1}.

\section{Review of classical A-Optimal Projection}
\label{sec2:level1}

In this section, we briefly review the AOP model and learning algorithm.

The classical AOP dimensionality reduction aims to improve the regression performance in the reduced space which preserves similarities between the data pairs. The AOP dimensionality reduction algorithm returns the directions of projections, and  with this result, the data can be projected onto a lower-dimensional subspace which can be directly used for regression problem.

Let ${\bf{X}} = \left( {{{\bf{x}}_1}, \cdots ,{{\bf{x}}_m}} \right)$ be a $n \times m$ data matrix, where $m$ is the number of data points and $n$ is the number of features. 
In the graph based dimensionality reduction, we are given a nearest neighbor graph $G$ which represents the geometrical structure of the data manifold. Each vertex of the graph represents a data point. Let ${\bf{S}} \in \mathbb R{^{m \times m}}$ be the weight matrix of the graph and $N_k\left({\bf{x}}\right)$ denote the $k$ nearest neighbors of ${\bf{x}}$. Then a simple example of ${\bf{S}}$ can be defined as follows:
                	\begin{equation}
                      {S_{ij}} = \left\{ \begin{array}{l}
				1, \quad if \quad {{\bf{x}}_i} \in {N_k}\left( {{{\bf{x}}_j}} \right) \quad 
					or \quad{{\bf{x}}_j} \in {N_k}\left( {{{\bf{x}}_i}} \right)\\
				0, \quad otherwise
				\end{array} \right.
                    \label{eq:z}
                	\end{equation}

AOP aims to find a projection matrix ${\bf{A}} \in \mathbb R{^{n \times k}}$ that maps the the points ${{\bf{x}}_i}$ to ${{\bf{y}}_i} \in \mathbb R{^k} $ $(i = 1,...,m$, and $k \ll n )$, where ${{\bf{y}}_i} = {{\bf{A}}^T}{{\bf{x}}_i}$. And using ${{\bf{y}}_i}$ to train a linear regression model:
                	\begin{equation}
                       z = {{\bf{\beta }}^T}{\bf{y}} + \epsilon_0 
                    \label{eq:z}
                	\end{equation}
where $z$ is the observation, ${\bf{\beta}}$ is the weight vector and $\epsilon_0$ is an unknown error with Gaussian distribution.

Formally, the objective function of AOP is:
                	\begin{equation}
                       \mathop {\min }\limits_{\bf{A}} 
                       Tr\left( {{{\left( {{{\bf{A}}^T}{\bf{X}}\left( {{\bf{I}} 
                       + {\lambda _1}{\bf{L}}} \right){{\bf{X}}^T}
                       {\bf{A}} + {\lambda _2}{\bf{I}}} \right)}^{ - 1}}} \right)
                    \label{eq:obj}
                	\end{equation}
where $ {\lambda _1}$ and $ {\lambda _2}$ are the regularization coefficients which are very small, and ${\bf{L}} = diag\left( {{\bf{S1}}} \right) - {\bf{S}}$ is the graph Laplacian (${\bf{1}}$ is a vector of all ones). 

To solve the objective function, Ref. \cite{HZZ16} introduces a variables {\bf{B}}  and the optimization problem (\ref{eq:obj}) is equivalent to the following:
                	\begin{equation}
                       \mathop {\min }\limits_{{\bf{A}},{\bf{B}}} {\left\| {{\bf{I}} - {{\bf{A}}^T}
                       \widetilde {\bf{X}}{\bf{B}}} \right\|^2} + {\lambda _2}
                       {\left\| {\bf{B}} \right\|^2}
                    \label{eq:obj1}
                	\end{equation}
where $\widetilde {\bf{X}} = {\bf{X}}\Sigma $,  and $\Sigma$ is from the cholesky decomposition: ${\bf{I}} + {\lambda _1}{\bf{L}} = \Sigma {\Sigma ^T}$.

It tells us that the optimal ${\bf{A}}$ can be obtained by iteratively computing ${\bf{A}}$ and ${\bf{B}}$. Then the overall procedure of the AOP learning algorithm is depicted as follows:

1) Initialize the matrix ${\bf{A}}$ by computing the PCA of the data ${\bf{X}}$.

2) Compute the matrix ${\bf{B}}$ according to the Eq. (\ref{eq:B}):

                	\begin{equation}
                       \frac{{\partial \phi }}{{\partial {{\bf{B}}^T}}} 
                       = 0 \Rightarrow {\bf{B}} 
                       = {\left( {{{\widetilde {\bf{X}}}^T}{\bf{A}}{{\bf{A}}^T}\widetilde {\bf{X}} 
                       + {\lambda _2}{\bf{I}}} \right)^{ - 1}}{\widetilde {\bf{X}}^T}{\bf{A}}
                    \label{eq:B}
                	\end{equation}

3) Update the matrix ${\bf{A}}$ according to the Eq. (\ref{eq:A}), and normalize ${\bf{A}}$ such that ${\left\| {\bf{A}} \right\|_F} \le {\rho_0} $.

                	\begin{equation}
                       \frac{{\partial \phi }}{{\partial {\bf{A}}}} 
                       = 0 \Rightarrow {\bf{A}} 
                       = {\left( {\widetilde {\bf{X}}{\bf{B}}{{\bf{B}}^T}
                       {{\widetilde {\bf{X}}}^T}} \right)^{ - 1}}\widetilde {\bf{X}}{\bf{B}}
                    \label{eq:A}
                	\end{equation}
where $\rho_0$ is used as a constraint parameter to control the size of ${\bf{A}}$.

4) Repeat steps 2 and 3 until convergence.

\section{Quantum A-Optimal Projection}
\label{sec3:level1}

In this section, we propose the quantum AOP algorithm for dimensionality reduction. We firstly reformulated the original classical AOP algorithm. And with the help of the reformulation, the proposed quantum AOP algorithm can then be implemented more efficiently.

\subsection {Reformulation of the AOP algorithm}
\label{sec3.1:level2}

We reformulated the algorithm in Sec. \ref{sec2:level1} in terms of quantum mechanics. 
Firstly, we adjust the initialization of ${\bf{A}}$ to make it closer to the optimal solution than the original algorithm. 
Secondly, we combine the steps 2 and 3 into one step and remove the variable {\bf{B}}. 
The advantage of eliminating {\bf{B}} is to help avoid quantum-classical transformation during the iteration of the algorithm.
Specifically, in one iteration of the original algorithm, a quantum state is needed to be computed and sampled as to reconstruct the matrix {\bf{B}}, and then used to update {\bf{A}}. 
In our methods, the elimination of the matrix {\bf{B}} can help update the quantum state representing {\bf{A}} without sampling and reconstruction in one of the iterations. 
Finally, by introducing the partial trace technology, quantum-classical interaction can also be omitted 
between the iterations.

(1) Initialization of ${\bf{A}}$. 
The original AOP algorithm compute PCA of the data ${\bf{X}}$ to initialize the matrix ${\bf{A}}$. In contrast, we compute PCA of ${\widetilde {\bf{X}}}$ to obtain the initialization ${{\bf{A}}^{\left( 0 \right)}}$. As shown in Eq. (\ref{eq:obj}), when $ {\lambda _1}$ and $ {\lambda _2}$ are set to zero, the objection function (\ref{eq:obj}) is equivalent to the objection function of PCA referring to the data ${\bf{X}}$.  And when $ {\lambda _2}$ is zero, the objection function (\ref{eq:obj}) is equivalent to the objection function of PCA referring to the data ${\widetilde {\bf{X}}}$. It is obvious that the later one is closer to the optimal solution than the former one. 

(2) Reformulation of the AOP algorithm. 
Now turning to the steps 2 and 3 of the AOP algorithm in one of the iterations. Suppose the singular value decomposition of the matrix ${\widetilde {\bf{X}}}$ is ${\widetilde {\bf{X}}}=\sum\nolimits_{j = 1}^r {{\sigma _j}\left| {{{\bf{u}}_j}} \right\rangle \left\langle {{{\bf{v}}_j}} \right|} $, 
where $r \le \min \left( {m,n} \right)$ is the rank of ${\widetilde {\bf{X}}}$, and ${\sigma _k} \left( {{\sigma _1} >  \cdots  > {\sigma _r} > 0} \right)$ are the singular values of ${\widetilde {\bf{X}}}$, with ${{\bf{u}}_j}$ and ${{\bf{v}}_j}$ being the left and right singular vectors. 
Obviously, we have
${\widetilde {\bf{X}}^T} = \sum\nolimits_{j = 1}^r {{\sigma _j}\left| {{{\bf{v}}_j}} \right\rangle \left\langle {{{\bf{u}}_j}} \right|} $ and $\widetilde {\bf{X}}{\widetilde {\bf{X}}^T} = \sum\nolimits_{j = 1}^r {\sigma _j^2\left| {{{\bf{u}}_j}} \right\rangle \left\langle {{{\bf{u}}_j}} \right|} $. 

As ${{\bf{A}}^{\left( 0 \right)}}$ is the PCA of ${\widetilde {\bf{X}}}$, we have 

	\begin{equation} 
		{{\bf{A}}^{\left( 0 \right)}} 
		= pca\left( {\widetilde {\bf{X}}} \right) 
		= \sum\limits_{j = 1}^k {\left| {{{\bf{u}}_j}} \right\rangle \left\langle {\bf{j}} \right|} ,
	\label{eq:A0}
        \end{equation} 
where $k$ is the rank of ${\bf{A}}$, and $\left|{\bf{j}}\right\rangle$'s are the basis states. 
Now we have the theorem 1 (and the proof is shown in Appendix  \ref{A1}.). 

\textbf{Theorem 1}:
 \emph{ Given the matrix ${\widetilde {\bf{X}}}=\sum\nolimits_{j = 1}^r {{\sigma _j}\left| {{{\bf{u}}_j}} \right\rangle \left\langle {{{\bf{v}}_j}} \right|} $, the $i$-th iteration of the AOP algorithm outputs the matrix ${{\bf{A}}^{\left( i \right)}}$:	
	\begin{equation}
		{{\bf{A}}^{\left( i \right)}} 
		=\sum\limits_{j = 1}^k { {\beta _j^{\left( i \right)}} \left| {{{\bf{u}}_j}} \right\rangle \left\langle {\bf{j}} \right|} 
		= \sum\limits_{j = 1}^k {
		\frac{{{{\left( {{\sigma _j}\beta _j^{\left( {i - 1} \right)}} \right)}^2} + {\lambda _2}}}
		        {{\sigma _j^2\beta _j^{\left( {i - 1} \right)}}}
		 \left| {{{\bf{u}}_j}} \right\rangle \left\langle {\bf{j}} \right|} 
	\label{eq:Anew}
        \end{equation}
where $1 \le i \le s$ and ${\beta _j^{\left( 0 \right)}}=1$ for all $j$'s.
}

According to the theorem 1, the reformulated AOP algorithm is presented as follows:

1) Initialize the matrix ${{\bf{A}}^{\left( 0 \right)}}$ by computing the PCA of ${\widetilde {\bf{X}}}$ according to Eq. (\ref{eq:A0}).

2) Update the matrix ${\bf{A}}$ according to Eq. (\ref{eq:Anew}).

3) Repeat step 2 until convergence.

\begin{figure}[ht]
\centering
\includegraphics[width=0.8\linewidth]{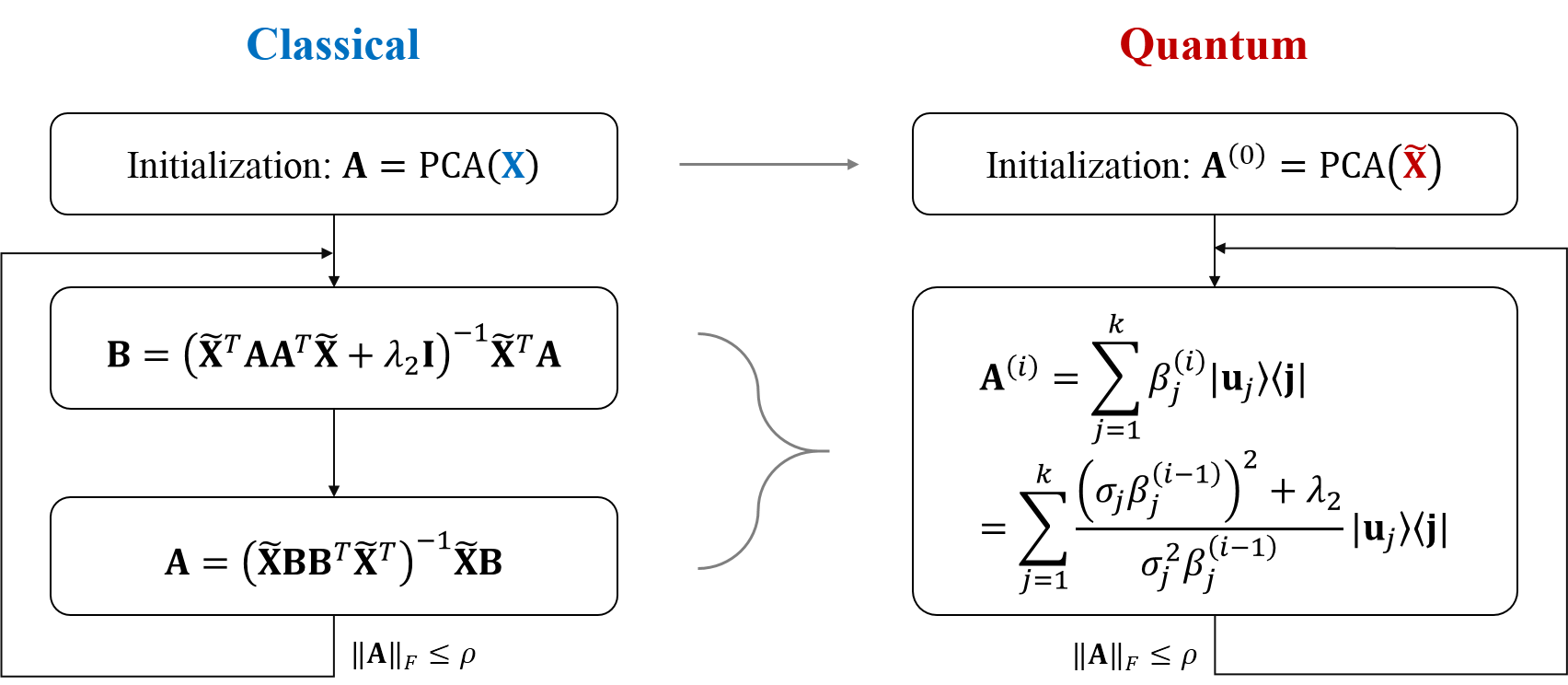}
\caption{Reformulation of the AOP algorithm.}
\label{fig1}
\end{figure}

The modeling of the AOP algorithm is shown in Fig. \ref{fig1}.

\subsection {QAOP algorithm }
\label{sec3.2:level2}

The overall procedure of our QAOP algorithm is then proposed as follows.

    \emph{Algorithm.} $\textbf{A}^{(s)}  = QAOP \left({\widetilde {\bf{X}}}, {\lambda _2}\right)$.

    1. Initialize $i=1$ , apply quantum PCA algorithm of ${\widetilde {\bf{X}}}$ to compute $\textbf{A}^{(0)}$ \cite{LMR14}.
   
    2. Perform one iteration $Q$ of the QAOP algorithm to compute $\textbf{A}^{(i)}$, i.e. $\textbf{A}^{(i)}  = Q \left( {\widetilde {\bf{X}}}, \textbf{A}^{(i-1)} \right)$.
    
    3. Set $i=i+1$, and repeat step 2 until the number of iterations $i = s$; now the projection matrix $\textbf{A}^{(s)}$ can be obtained.

The quantum algorithm for $i$-th iteration $Q$ of the QAOP algorithm in Step 2 is presented as follows.

     (1) Prepare four quantum registers in the state
        \begin{equation}
        \begin{aligned}
            \left| {{\psi _0 }} \right\rangle  = \left| 0 \right\rangle^a 
            \left( {\left| 0 \right\rangle \left| 0 \right\rangle  \cdots \left| 0 \right\rangle } \right)^{C} 
            \left( {\left| 0 \right\rangle \left| 0 \right\rangle  \cdots \left| 0 \right\rangle } \right)^{B} 
            \left( {\left| {{\psi _{\textbf{A}^{(i-1)}}}} \right\rangle } \right)^A.
        \end{aligned}
        \end{equation}
where the superscript $a$ represents the ancilla qubit, the superscripts ${C}, {B}, A$ represent the register ${C}, {B}$ and $A$, respectively. 

     (2) Perform the unitary operation ${{\bf{U}}_{PE}}\left( {\widetilde {\bf{X}}} {\widetilde {\bf{X}}}^\dag \right)$ and ${{\bf{U}}_{PE}}\left( {\textbf{A}^{(i-1)}} {\textbf{A}^{(i-1)}}^\dag \right)$  on the state, then we have the state
        \begin{equation}
        \begin{aligned}
            \left| {{\psi _1}} \right\rangle  
            = \frac{1}{{\sqrt {{N_1}} }}{\left| 0 \right\rangle ^a}\sum\limits_{j = 1}^k 
            {\beta _j^{\left( {i - 1} \right)}{{\left| {\sigma _j^2} \right\rangle }^{C}}
            {{\left| {{{\left( {\beta _j^{\left( {i - 1} \right)}} \right)}^2}} \right\rangle }^{B}}
            \left| {{{\bf{u}}_j}} \right\rangle {{\left| {{{\bf{v}}_j}} \right\rangle }^A}} .
        \end{aligned}
        \end{equation}
Here ${{\bf{U}}_{PE}}$ represents the unitary matrix for phase estimation which we fully characterized in \cite{Duan17}:

\begin{equation}
        \begin{aligned}
            {{\bf{U}}_{PE}}\left( {\bf{X}} \right) = \left( {{\bf{F}}_{{\bf{T}}}^\dag \otimes {{\bf{I}}}} \right)\left( {\sum\nolimits_{\tau  = 0}^{T - 1} {\left| \tau  \right\rangle {{\left\langle \tau  \right|}} \otimes {e^{i{\bf{X}}\tau {{{t_0}} \mathord{\left/  {\vphantom {{{t_0}} T}} \right. \kern-\nulldelimiterspace} T}}}} } \right)\left( {{{\bf{H}}^{ \otimes t}} \otimes {{\bf{I}}}} \right),
        \end{aligned}
        \label{eq:U_PE}
		\end{equation}
    where  ${\bf{F}}_{{\bf{T}}}^\dag$ is the inverse quantum Fourier transform and ${\sum\nolimits_{\tau  = 0}^{T - 1} {\left| \tau  \right\rangle {{\left\langle \tau  \right|}^C} \otimes {e^{i{\bf{A}}\tau {{{t_0}} \mathord{\left/ {\vphantom {{{t_0}} T}} \right. \kern-\nulldelimiterspace} T}}}} }$ is the conditional Hamiltonian evolution \cite{HHL09}.

     (3) Apply a controlled rotation ${\bf{R}}_{f}$ to the ancilla qubit, controlled by both the register ${C}$ and ${B}$. ${\bf{R}}_{f}$ is defined as follows:
        \begin{equation}
        \begin{aligned}
		{R_f}: &{\left| 0 \right\rangle ^a}{\left| {\sigma _j^2} \right\rangle ^{C}}
        		{\left| {{{\left( {\beta _j^{\left( {i - 1} \right)}} \right)}^2}} \right\rangle ^{B}}\\
		& \to {\left[ {\rho \left( {1 + \frac{{{\lambda _2}}}
		{{{{\left( {{\sigma _j}\beta _j^{\left( {i - 1} \right)}}\right)}^2}}}} \right)\left| 1 \right\rangle  
		+ \sqrt {1 - {\rho ^2}{{\left( {1 + \frac{{{\lambda _2}}}
		{{{{\left( {{\sigma _j}\beta _j^{\left( {i - 1} \right)}} \right)}^2}}}} \right)}^2}} \left| 0 \right\rangle } \right]^a}
		{\left| {\sigma _j^2} \right\rangle ^{C}}{\left| {{{\left( {\beta _j^{\left( {i - 1} \right)}} \right)}^2}} \right\rangle ^{B}},
        \end{aligned}
        \label{eq:Rf}
        \end{equation}
where $\rho<1$ is a parameter used for normalization of the quantum state.

     This rotation transforms the state to
            \begin{equation}
            \begin{aligned}
                \left| {{\psi _2}} \right\rangle  
                = & \frac{1}{{\sqrt {{N_1}} }} {{\left[ {\rho \left( {1 + \frac{{{\lambda _2}}}
                {{{{\left( {{\sigma _j}\beta _j^{\left( {i - 1} \right)}} \right)}^2}}}} \right)\left| 1 \right\rangle  
                + \sqrt {1 - {\rho ^2}{{\left( {1 + \frac{{{\lambda _2}}}
                {{{{\left( {{\sigma _j}\beta _j^{\left( {i - 1} \right)}} \right)}^2}}}} \right)}^2}}
                 \left| 0 \right\rangle } \right]}^a} \\
                & \otimes \sum\limits_{j = 1}^k {\beta _j^{\left( {i - 1} \right)}
                 {{\left| {\sigma _j^2} \right\rangle }^{C}}
                 {{\left| {{{\left( {\beta _j^{\left( {i - 1} \right)}} \right)}^2}} \right\rangle }^{B}}
                 \left| {{{\bf{u}}_j}} \right\rangle {{\left| {{{\bf{v}}_j}} \right\rangle }^A}} .
            \end{aligned}
            \end{equation}

    (4) Uncompute the registers $C$, $B$ and $A$, remove the register $C$ and $B$, and measure the ancilla qubit to be $\left| 1 \right\rangle$. Then, we have the state proportional to
            \begin{equation}
            \begin{aligned}
               	\left| {\psi _{\bf{A}}^{\left( i \right)}} \right\rangle  
		= \frac{1}{{\sqrt {{N_2}} }}\sum\limits_{j = 1}^k 
		{\frac{{{{\left( {{\sigma _j}\beta _j^{\left( {i - 1} \right)}} \right)}^2} 
		+ {\lambda _2}}}{{\sigma _j^2\beta _j^{\left( {i - 1} \right)}}}
		\left| {{{\bf{u}}_j}} \right\rangle {{\left| {{{\bf{v}}_j}} \right\rangle }^A}} .
            \end{aligned}
            \end{equation}

 	Here, we can construct the matrix ${\textbf{A}^{(i)}} {\textbf{A}^{(i)}}^\dag$ for the phase estimation in the next iteration by taking a partial trace of $\left| {\psi _{\bf{A}}^{\left( i \right)}} \right\rangle \left\langle {\psi _{\bf{A}}^{\left( i \right)}} \right|$. Note that 
the eigenvectors of ${\textbf{A}^{(i)}} {\textbf{A}^{(i)}}^\dag$ are ${{\bf{u}}_j}$ and the corresponding eigenvalues are ${\left( {\beta _j^{\left( {i} \right)}} \right)}^2$. Then the density matrix that represents ${\textbf{A}^{(i)}} {\textbf{A}^{(i)}}^\dag$ can be obtained \cite{SSP16} : 

            \begin{equation}
            \begin{aligned}
		t{r_2}\left( \left| {\psi _{\bf{A}}^{\left( i \right)}} \right\rangle \left\langle {\psi _{\bf{A}}^{\left( i \right)}} \right| \right)
                = \frac{1}{\sum\nolimits_{k = 1}^r {{\left( {\beta _j^{\left( {i} \right)}} \right)}^2}}
                \sum\nolimits_{k = 1}^r {{\left( {\beta _j^{\left( {i} \right)}} \right)}^2 \left| {{{\bf{u}}_k}} \right\rangle 
                \left\langle {{{\bf{u}}_k}} \right| 
                = \frac{{\textbf{A}^{(i)}} {\textbf{A}^{(i)}}^\dag} 
                {tr \left( {{\textbf{A}^{(i)}} {\textbf{A}^{(i)}}^\dag} \right) }}.
            \end{aligned}
            \end{equation}

\section {QAOP circuit}
\label{sec4:level1}

In this section, we study the QAOP algorithm in terms of the quantum circuit model. The quantum circuits provide the possibility to implement the quantum algorithm on a universal quantum computer.
First, we present the overview model of the QAOP circuit. Second, we study in depth the realization of the initial state preparation and the controlled rotation in terms of the quantum circuit model. Finally, the space and time resources required for the quantum circuit are analyzed.

\begin{figure}[ht]
\centering
\includegraphics[width=0.9\linewidth]{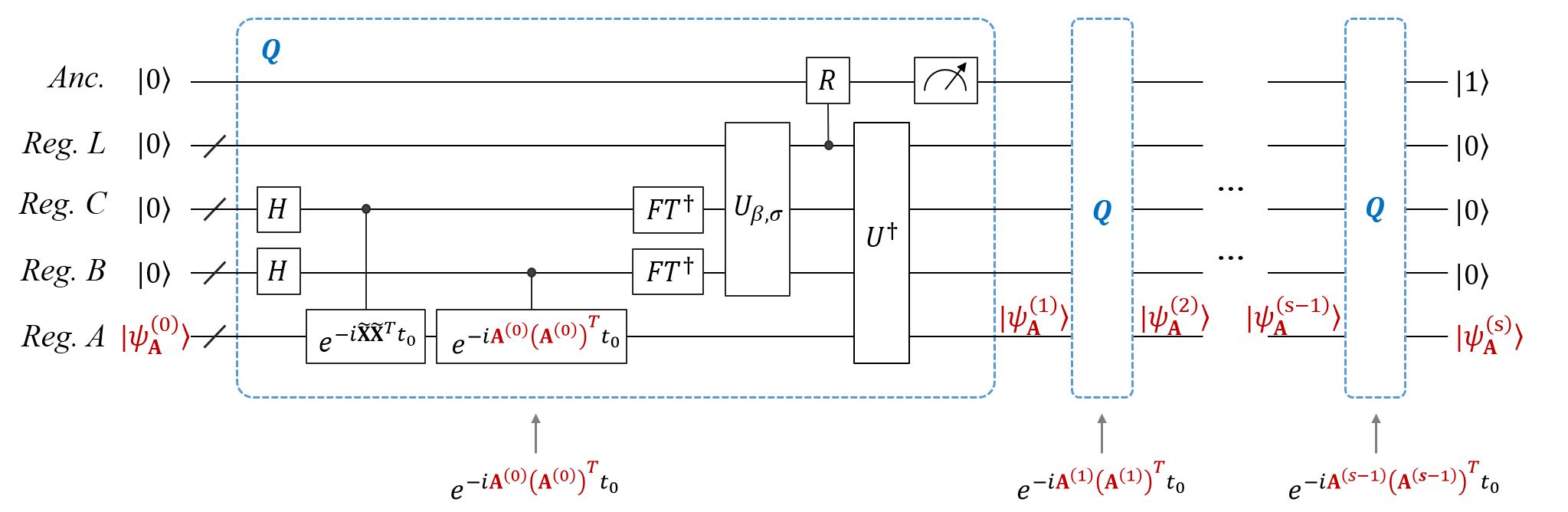}
\caption{Overview of the quantum circuit for solving the reformulated AOP. 
Wires with '/' represent the groups of qubits. The label Q in the dotted box represents one iteration of QAOP algorithm. }
\label{fig2}
\end{figure}

The overview of the circuit for solving QAOP is shown in Fig. \ref{fig2}.
It provides one model to implement the QAOP algorithm. Take the $i$-th iteration of the QAOP algorithm for example. It can be divided into three major steps: 
(1) Phase estimation: as the eigenspace of the unitary ${e^{ - i\widetilde {\bf{X}}{{\widetilde {\bf{X}}}^T}{t_0}}}$ and ${e^{ - i{{\bf{A}}^{\left( {i - 1} \right)}}{{\left( {{{\bf{A}}^{\left( {i - 1} \right)}}} \right)}^T}{t_0}}}$ are both spanned by the eigenvectors ${\left| {{{\bf{u}}_j}} \right\rangle }$, they can be both implemented on the input state ${\left| {\psi _{\bf{A}}^{\left( {i - 1} \right)}} \right\rangle }$. And the Hadamard gates and the inverse QFT of ${{\bf{U}}_{PE}}\left( {\widetilde {\bf{X}}} {\widetilde {\bf{X}}}^\dag \right)$ and ${{\bf{U}}_{PE}}\left( {\textbf{A}^{(i-1)}} {\textbf{A}^{(i-1)}}^\dag \right)$ can be implemented in parallel. 
(2) Controlled rotation: it consists of ${U_{\beta ,\sigma }}$ and $c-R_y$. Firstly the operation ${U_{\beta ,\sigma }}$ computes the function ${y_j}$ of the output eigenvalues of ${\widetilde {\bf{X}}} {\widetilde {\bf{X}}}^\dag $ and ${\textbf{A}^{(i-1)}} {\textbf{A}^{(i-1)}}^\dag $ as follows:

            \begin{equation}
            \begin{aligned}
               	{y_j} = 1 + \frac{{{\lambda _2}}}{{{{\left( {{\sigma _j}\beta _j^{\left( {i - 1} \right)}} \right)}^2}}} .
            \end{aligned}
            \end{equation}
And then the operation $c-R_y$ extracts the value of $y_j$ in the basic states of the register $L$ to the amplitude of 
the ancilla qubit. 
(3)  Uncomputing: undo the Reg. $B,C$ and $L$, and measure the top ancilla qubit. If the result returns to 1, then the Reg. $A$ of the quantum system collapses to the output state ${\left| {\psi _{\bf{A}}^{\left( {i} \right)}} \right\rangle }$, which is also the input state of the $\left(i+1 \right)$-th iteration.

We now deal with the detailed QAOP circuit. In the following, we mainly investigate the quantum circuits for the initial state preparation and the controlled rotation in one iteration of the QAOP circuit.

\subsection {State preparation}
\label{sec4.1:level2}	
\begin{figure}[ht]
\centering
\includegraphics[width=1\linewidth]{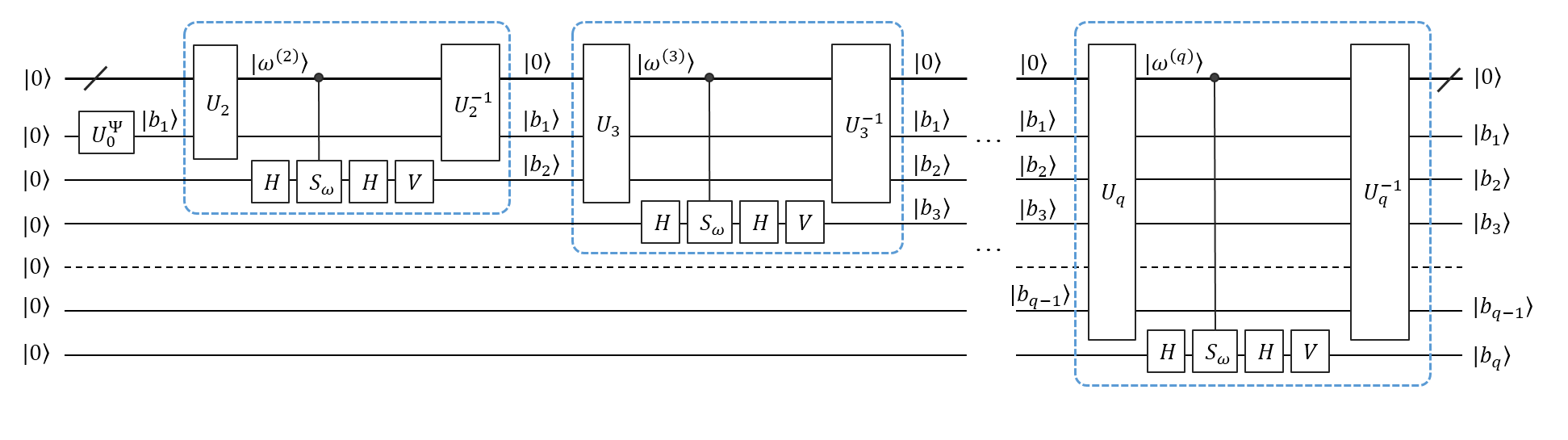}
\caption{The circuit for state preparation. }
\label{fig3}
\end{figure}

At the very beginning, we present a detailed quantum circuit for preparing the initial state of the QAOP algorithm. Suppose each element of ${\textbf{A}^{(0)}} \in \mathbb R{^{n \times k}}$ is given, and its corresponding quantum state ${\left| {{\psi _{\textbf{A}^{(0)}}}} \right\rangle } = \left| {{a_1}{a_2} \cdots {a_{q}}} \right\rangle$ is a $q$-qubit quantum state, where $q=O\left [ \log_2 \left ( nk \right )\right ]$. 
Following the approach in \cite{KM02}, the quantum circuit for the initial state preparation can be shown in Fig. \ref{fig3}. Here, we introduce quantum random access memory (QRAM) to omit the register $\left| {\bar \psi } \right\rangle $ in \cite{KM02}.

In Fig. \ref{fig3}, an register of $p=O\left (\log_2 \epsilon ^{-1}\right )$ qubits is used for storing the ${\omega ^{\left( i \right)}}$, where ${\omega ^{\left( i \right)}}$ is short for $\omega _{{a_1} \cdots {a_{i - 1}}} \left(i=1,2,3,...,q \right)$ satisfying:
            \begin{equation}
          	  {\cos ^2}\left( {2\pi \omega _{{a_1} \cdots {a_{i - 1}}}} \right) 
          	  = {\left( {\frac{{{\alpha _{{a_1}{a_2} \cdots {a_{i - 1}}0}}}}
           	 {{{\alpha _{{a_1}{a_2} \cdots {a_{i - 1}}}}}}} \right)^2} + 
           	 O\left( {poly\left( \epsilon  \right)} \right).
            \end{equation} 
All the $\omega _{{a_1} \cdots {a_{i - 1}}}$ can be computed classically and they are supposed to be stored in the QRAM. Given the index ${{a_1}{a_2} \cdots {a_{i - 1}}}$, define $j=h\left({{a_1}{a_2} \cdots {a_{i - 1}}}\right)$ being the address where the data ${\omega _{{a_1}{a_2} \cdots {a_{i - 1}}}}$ stores, where $h\left(  \cdot  \right)$ is a hash function mapping ${{a_1}{a_2} \cdots {a_{i - 1}}}$ to $j$.
Then define the unitary operation $U_i$ which implements the QRAM readout operation \cite{GLM08a}:

          \begin{equation}   
          \begin{aligned}      	  
		{U_i}: &\sum\limits_{{a_1},{a_2}, \cdots ,{a_{i - 1}} \in \{{0,1}\} } 
          	{{\alpha _{{a_1}{a_2} \cdots {a_{i - 1}}}}
		\left| {{a_1}{a_2} \cdots {a_{i - 1}}} \right\rangle } \left| j \right\rangle  
		\left| 0 \right\rangle\\
         	 \overset{QRAM}{\rightarrow} 
	 	 &\sum\limits_{{a_1},{a_2}, \cdots ,{a_{i - 1}} \in  \{{0,1}\}} 
         	 {{\alpha _{{a_1}{a_2} \cdots {a_{i - 1}}}}\left| {{a_1}{a_2} \cdots {a_{i - 1}}} \right\rangle 
	 	\left| j \right\rangle 
		\left| {{\omega _{{a_1}{a_2} \cdots {a_{i - 1}}}}}  \right\rangle } .
	 \end{aligned}
	 \end{equation}
specifically, $U_i$ outputs the content ${\omega _{{a_1}{a_2} \cdots {a_{i - 1}}}}$ of the $j$-th memory cell in QRAM.  Ref. \cite{HXZ12} shows that this procedure can be implemented in time $O\left(p \right)$. 

And the number of ${U_i}$ in the circuit for state preparation is $q$, so the total memory calls of QRAM is $O\left(pq\right)$.
Therefore, an inverse in $pq$ error rate suffices to achieve an overall constant error per QRAM look-up \cite{AGJ15}.

\begin{figure}[ht]
\centering
\includegraphics[width=0.8\linewidth]{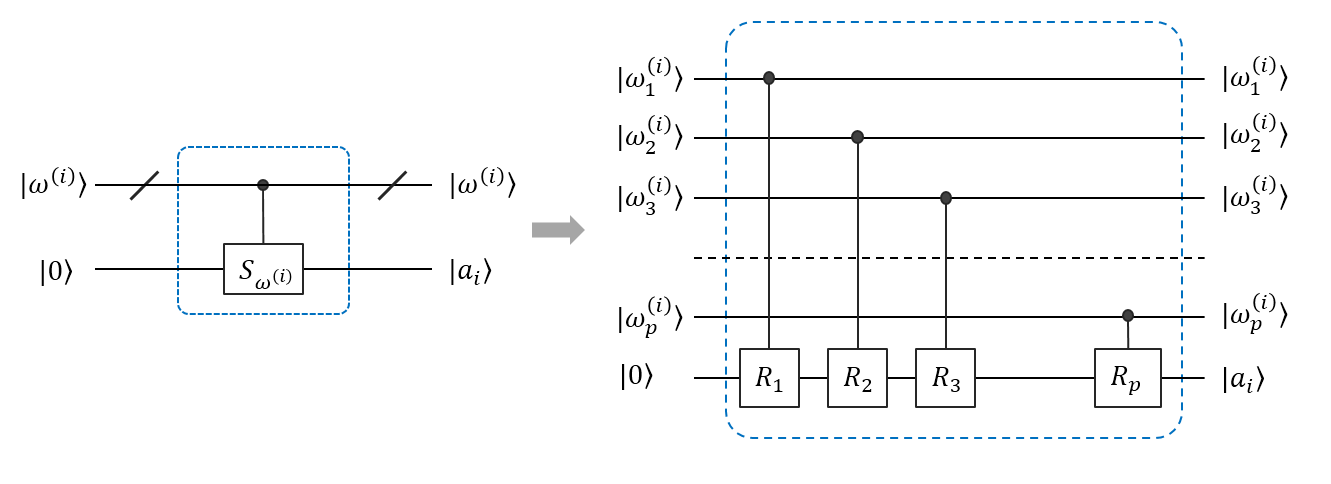}
\caption{The circuit for $c - {S_{\omega^{\left( i \right)}}}$, where 
$\left| {{a_i}} \right\rangle  
= \frac{{{\alpha _{{a_1} \cdots {a_{i - 1}}0}}}}{{{\alpha _{{a_1} \cdots {a_{i - 1}}}}}}\left| 0 \right\rangle  
+ \frac{{{\alpha _{{a_1} \cdots {a_{i - 1}}1}}}}{{{\alpha _{{a_1} \cdots {a_{i - 1}}}}}}\left| 1 \right\rangle $. }
\label{fig4}
\end{figure}

Moreover, we make further study on $c - {S_{\omega^{\left( i \right)}}}$ which is defined in Ref. \cite{KM02}: 
          \begin{equation}   
          \begin{aligned}  
		c - {S_{\omega^{\left( i \right)}}}:\left| \omega ^{\left( i \right)} \right\rangle 
	  	\left| 0 \right\rangle  \to \left| \omega ^{\left( i \right)} \right\rangle 
	  	{e^{2\pi i\omega ^{\left( i \right)}}}\left| 0 \right\rangle ,
		\left| \omega ^{\left( i \right)} \right\rangle \left| 1 \right\rangle  
		\to \left| \omega ^{\left( i \right)} \right\rangle 
		{e^{ - 2\pi i\omega^{\left( i \right)} }}\left| 1 \right\rangle .
	 \end{aligned}
	 \end{equation}			 
Define ${R_l} = \left( {\begin{array}{*{20}{c}}
		{{e^{{{2\pi i} \mathord{\left/
		 {\vphantom {{2\pi i} {{2^l}}}} \right.
 		\kern-\nulldelimiterspace} {{2^l}}}}}}&0\\
		0&{{e^{{{ - 2\pi i} \mathord{\left/
 		{\vphantom {{ - 2\pi i} {{2^l}}}} \right.
 		\kern-\nulldelimiterspace} {{2^l}}}}}}
		\end{array}} \right)$,
then the controlled unitary $c-R_l$ implements the following transformation:

	\begin{equation}
	\begin{aligned}
		\left| {\omega _l^{\left( i \right)}} \right\rangle 
		\left| 0 \right\rangle  
		&\to {e^{2\pi i{{\omega _l^{\left( i \right)}} 
		\mathord{\left/
 		{\vphantom {{\omega _l^{\left( i \right)}} {{2^l}}}} \right.
 		\kern-\nulldelimiterspace} {{2^l}}}}}\left| {\omega _l^{\left( i \right)}} \right\rangle 
		\left| 0 \right\rangle \\
		\left| {\omega _l^{\left( i \right)}} \right\rangle 
		\left| 1 \right\rangle  
		&\to {e^{ - 2\pi i{{\omega _l^{\left( i \right)}} \mathord{\left/
		 {\vphantom {{\omega _l^{\left( i \right)}} {{2^l}}}} \right.
 		\kern-\nulldelimiterspace} {{2^l}}}}}
		\left| {\omega _l^{\left( i \right)}} \right\rangle \left| 1 \right\rangle 
	\end{aligned}
        \end{equation}
where $\omega_l^{\left( i \right)}$ is the $l$-th binary bit of $\omega^{\left( i \right)}$, specifically,  ${\omega ^{\left( i \right)}} = {2^{ - 1}}\omega _1^{\left( i \right)} + {2^{ - 2}}\omega _2^{\left( i \right)} +  \cdots  + {2^{ - p}}\omega _p^{\left( i \right)} = 0.\omega _1^{\left( i \right)}\omega _2^{\left( i \right)} \cdots \omega _p^{\left( i \right)}$.
Now $\prod\limits_{l = 1}^p {\left( {c - {R_l}} \right)} $ achieves the function of $c - {S_{\omega^{\left( i \right)}}}$:

	\begin{equation}
	\begin{aligned}
		\left| {\omega _1^{\left( i \right)}} \right\rangle 
		\left| {\omega _2^{\left( i \right)}} \right\rangle  			
		\cdots \left| {\omega _p^{\left( i \right)}} \right\rangle 
		\left| 0 \right\rangle  
		&\to {e^{i2\pi 
		\left( {0.\omega _1^{\left( i \right)}
		\omega _2^{\left( i \right)} \cdots \omega _p^{\left( i \right)}} \right)}}
		\left| {\omega _1^{\left( i \right)}} \right\rangle 
		\left| {\omega _2^{\left( i \right)}} \right\rangle  
		\cdots \left| {\omega _p^{\left( i \right)}} \right\rangle \left| 0 \right\rangle \\
		\left| {\omega _1^{\left( i \right)}} \right\rangle 
		\left| {\omega _2^{\left( i \right)}} \right\rangle  
		\cdots \left| {\omega _p^{\left( i \right)}} \right\rangle 
		\left| 1 \right\rangle  
		&\to {e^{ - i2\pi 
		\left( {0.\omega _1^{\left( i \right)}\omega _2^{\left( i \right)}
		\cdots \omega _p^{\left( i \right)}} \right)}}
		\left| {\omega _1^{\left( i \right)}} \right\rangle 
		\left| {\omega _2^{\left( i \right)}} \right\rangle  
		\cdots \left| {\omega _p^{\left( i \right)}} \right\rangle 
		\left| 1 \right\rangle 
		\label{eq_cRk}
	\end{aligned}
         \end{equation}	
Therefore, the quantum circuit for $c - {S_{\omega^{\left( i \right)}}}$ can be implemented as shown in Fig. \ref{fig4}. 

Now we can simply infer that the number of qubits needed for preparing the initial quantum state ${\left| {{\psi _{\textbf{A}^{(0)}}}} \right\rangle }$ is $O\left(p+q \right)$, and the number of gates required is $O\left ( pq \right )$. 

In summary, with $V = \left( {\begin{array}{*{20}{c}}1&0\\ 0&{ - \iota}\end{array}} \right)$, the unitary $\left( {I \otimes V} \right)\left( {I \otimes H} \right)\left( {c - {S_\omega}} \right)\left( {I \otimes H} \right)$ implements the transformation:
	\begin{equation}
	\begin{aligned}
		&\left| \omega ^{\left( i \right)} \right\rangle \left| 0 \right\rangle \\ 
		\overset{I \otimes H}{\rightarrow}
		&\left| \omega ^{\left( i \right)} \right\rangle 
		\frac{{\left| 0 \right\rangle  + \left| 1 \right\rangle }}{{\sqrt 2 }}\\
		\overset{c - {S_\omega}}{\rightarrow}
		&\left| \omega  ^{\left( i \right)}\right\rangle \frac{{{e^{2\pi \iota \omega^{\left( i \right)} }}
		\left| 0 \right\rangle  + {e^{ - 2\pi \iota \omega ^{\left( i \right)}}}			
		\left| 1 \right\rangle }}{{\sqrt 2 }}\\
		\overset{I \otimes H}{\rightarrow}
		&\left| \omega ^{\left( i \right)} \right\rangle \left[ {\cos \left( {2\pi \omega^{\left( i \right)} } \right)
		\left| 0 \right\rangle  + \iota \sin 		
		\left( {2\pi \omega ^{\left( i \right)}} \right)\left| 1 \right\rangle } \right]\\
		\overset{I \otimes V}{\rightarrow}
		&\left| \omega ^{\left( i \right)} \right\rangle \left[ {\cos \left( {2\pi \omega ^{\left( i \right)}} \right)
		\left| 0 \right\rangle  + \sin \left( {2\pi \omega ^{\left( i \right)}} \right)\left| 1 \right\rangle } \right] \\
		=&\left| {{\omega ^{\left( i \right)}}} \right\rangle \left| {{a_i}} \right\rangle 
	\end{aligned}
         \end{equation}

\subsection {The controlled rotation}
\label{sec4.2:level2}	
The controlled rotation mainly involves the computation of $U_{\beta,\sigma}$ and $c-R_y$. In the stage of  $U_{\beta,\sigma}$, Newton's method is introduced for computing $y_j=y\left( {\sigma _j^2,\beta _j^2} \right) =  \frac{\rho \left({\sigma _j^2\beta _j^2 + {\lambda _2}} \right)}{{ \sigma _j^2\beta _j^2}} = \rho  + \frac{{\rho {\lambda _2}}}{{\sigma _j^2\beta _j^2}}$,	
where $\rho<1$ is used for normalization of the quantum state.
The value of $y_j$ are computed out and stored in the basis state of the register $L$, and the number of qubits for storing $y_j$ is $d = O\left( {{{\log }_2}\kappa } \right)$. In the stage of $c-R_y$,  $y_j$ is used as controlled qubit controlling the top ancilla qubit in Fig. \ref{fig2}.

\begin{figure}[ht]
\centering
\includegraphics[width=0.8\linewidth]{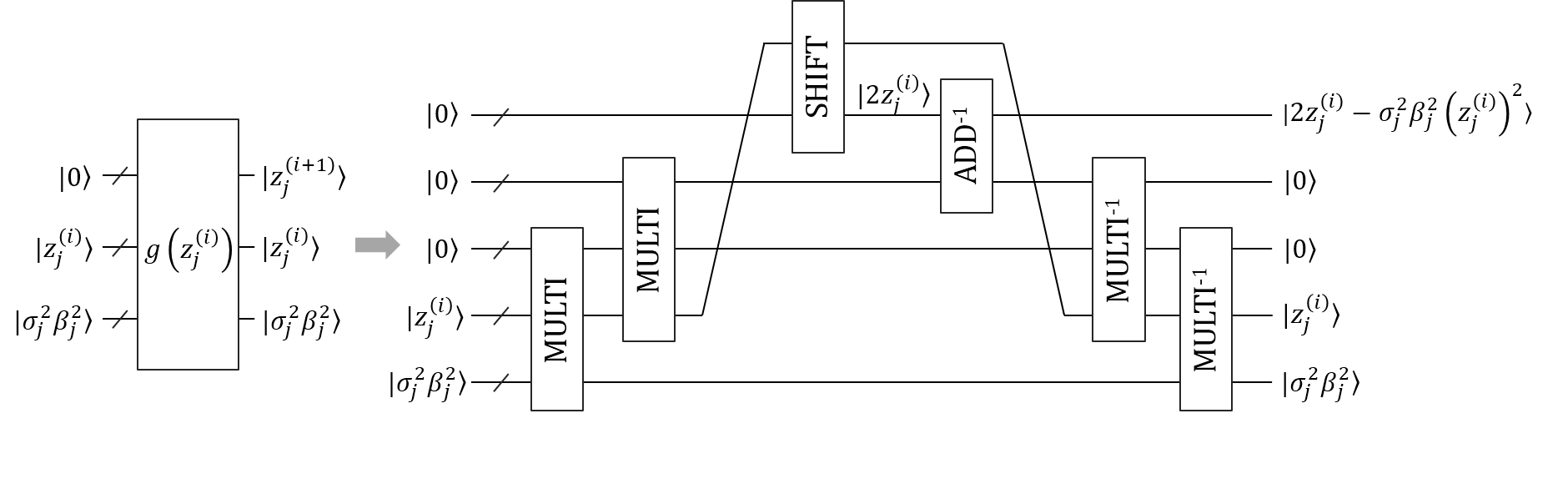}
\caption{The circuit for $z_j^{\left( {i + 1} \right)} =  - \sigma _j^2\beta _j^2{\left( {z_j^{\left( i \right)}} \right)^2} + 2z_j^{\left( i \right)}$. }
\label{fig5}
\end{figure}

(1) For $U_{\beta,\sigma}$, let ${z_j} = z\left( {\sigma _j^2\beta _j^2} \right) = {1 \mathord{\left/ {\vphantom {1 {\left( {\sigma _j^2\beta _j^2} \right)}}} \right. \kern-\nulldelimiterspace} {\left( {\sigma _j^2\beta _j^2} \right)}}$, 
then we have ${y_j} = \rho  + \rho {\lambda _2} z_j^{\left( {s'} \right)}$. 
Here we use Newton iteration to approximate ${1 \mathord{\left/  {\vphantom {1 {\left( {\sigma _j^2\beta _j^2} \right)}}} \right. \kern-\nulldelimiterspace} {\left( {\sigma _j^2\beta _j^2} \right)}}$, for ${\sigma _j^2\beta _j^2}>1$.
The quantum circuit for computing the initial approximation ${z_j^{\left( 0 \right)}}$ can be seen in \cite{CPP12}.
Applying the Newton method to  
$f\left( {{z_j}} \right) = {1 \mathord{\left/ {\vphantom {1 {{z_j}}}} \right. \kern-\nulldelimiterspace} {{z_j}}} - \sigma _j^2\beta _j^2$, we can get the Newton iteration function:
$z_j^{\left( {i + 1} \right)} = g\left( {z_j^{\left( i \right)}} \right) = z_j^{\left( i \right)} - \frac{{f\left( {z_j^{\left( i \right)}} \right)}}{{f'\left( {z_j^{\left( i \right)}} \right)}} =  - \sigma _j^2\beta _j^2{\left( {z_j^{\left( i \right)}} \right)^2} + 2z_j^{\left( i \right)}$.
Then the detailed circuit for one iteration of Newton's method is presented in Fig. \ref{fig5}. 
The number of qubits needed in the ancilla registers is $O(d)$. The number of fundamental quantum operations for implementing addition and multiplication is $O\left[poly \left(d \right) \right]$, where the degree of the polynomial is no more than $3$ \cite{VBE96,PRM17}, and the number of gates for implementing shift is $O(d)$.

\begin{figure}[ht]
\centering
\includegraphics[width=0.8\linewidth]{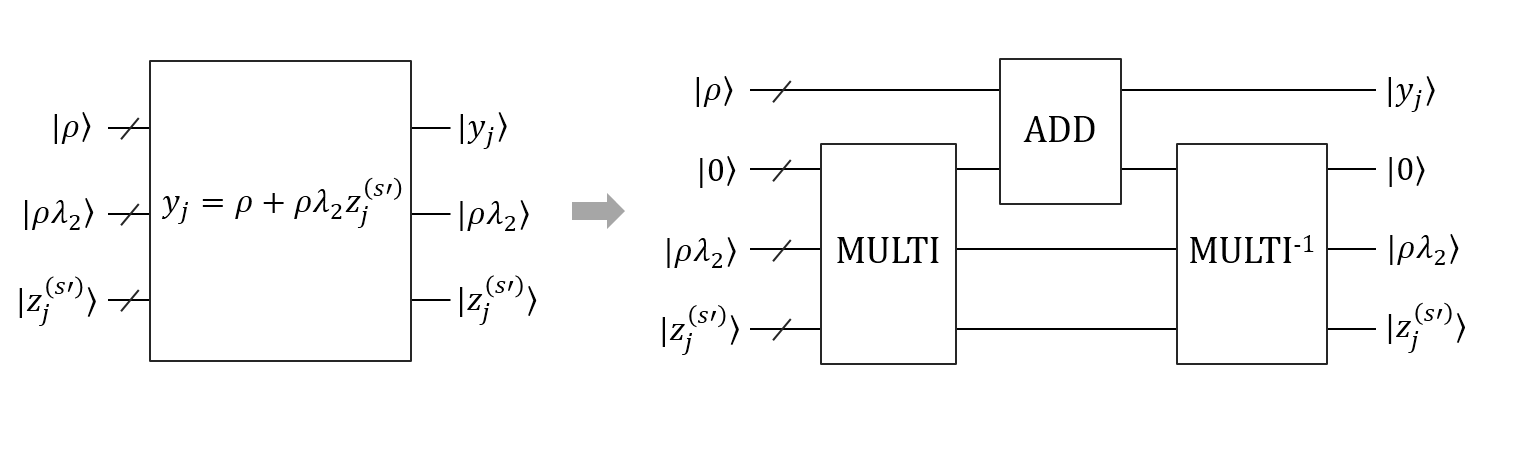}
\caption{The circuit for ${y_j} $. }
\label{fig6}
\end{figure}

The quantum circuit for ${y_j} = \rho  + \rho {\lambda _2} z_j^{\left( {s'} \right)}$ can be simply realized as shown in Fig. \ref{fig6}. 
The circuit can be simply realized with the quantum circuits for addition and multiplication. Therefore, the number of qubits and gates needed in the circuit are $O(d)$ and $O\left[ poly \left(d \right) \right]$, respectively.

\begin{figure}[ht]
\centering
\includegraphics[width=1\linewidth]{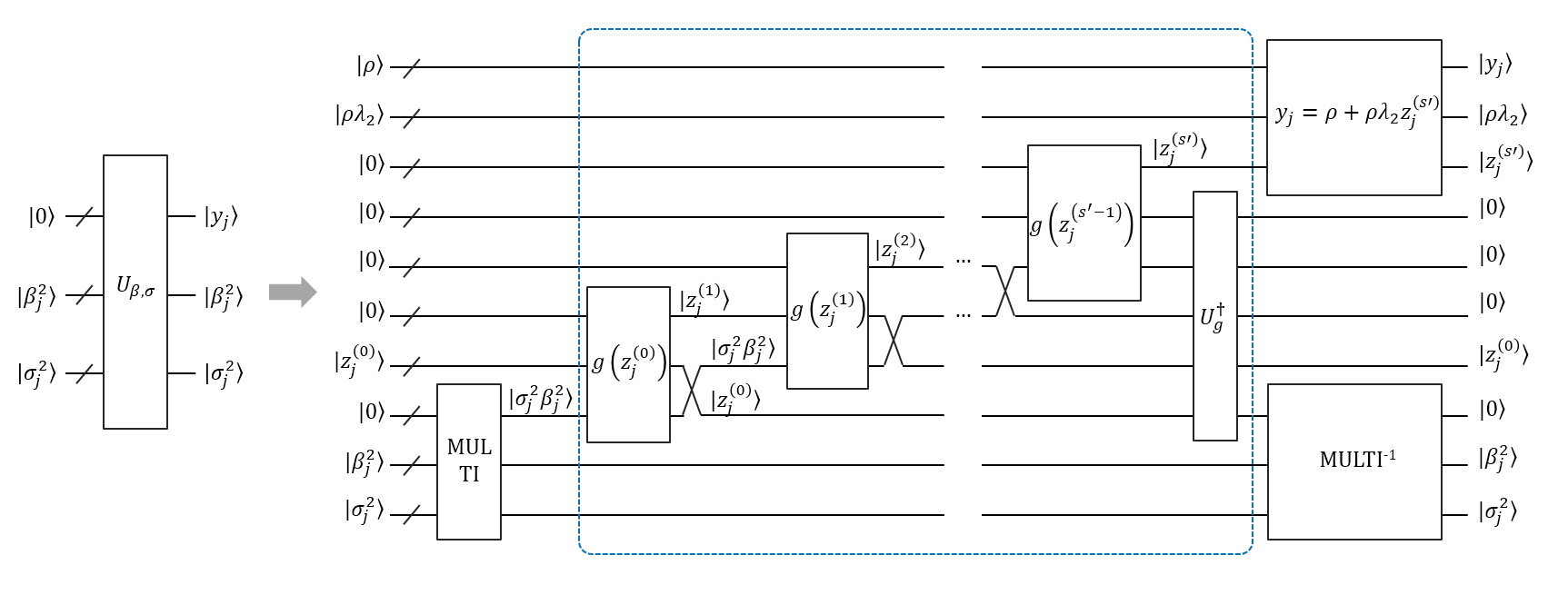}
\caption{The circuit for the unitary $U_{\beta,\sigma}$. }
\label{fig7}
\end{figure}

To sum up, the overall quantum circuit for $U_{\beta,\sigma}$ can be designed as shown in Fig. \ref{fig7}. 
And we can simply infer that the number of qubits needed in these circuits is $O(d+b)$. Let the number of Newton iteration be $s'$, then the number of gates required in Fig. \ref{fig7} is $O\left[s'poly \left(d \right) \right] $.

        Now we analyze the error caused by Newton's iteration. Similar to the error analysis in \cite{Duan18,CPP12}, the error consists of two parts. One is error $e_{s'}$ caused by the Newton's iteration, the other is the roundoff error ${{\hat e}_{s'}}$ caused by truncating the result of one iteration to $d$ qubits of accuracy. 
   
	According to the Newton iteration function, we have $g\left( {z_j^{\left( i \right)}} \right) -\frac{1}{{\sigma _j^2\beta _j^2}}  
=  - \sigma _j^2\beta _j^2{\left( {z_j^{\left( i \right)} - \frac{1}{{\sigma _j^2\beta _j^2}}} \right)^2}$.         
Then the error $e_{s'}$ satisfies          
${e_{s'}}:= \left| {z_j^{\left( {s'} \right)} - \frac{1}{{\sigma _j^2\beta _j^2}}} \right| = \sigma _j^2\beta _j^2 e_{{s'} - 1}^2
= \frac{1}{{\sigma _j^2\beta _j^2}}{\left( {\sigma _j^2\beta _j^2{e_0}} \right)^{{2^{s'}}}}$.                     
Following the approach in  \cite{CPP12}, the initial error ${e_0}$  satisfies  ${\sigma _j^2\beta _j^2} {e_0} < 1/2$, then for error ${\varepsilon _N}$ we have ${2^{ - {2^{s'}}}} \le {\varepsilon _N}$, which implies ${s'} \ge \left\lceil {{{\log }_2}{{\log }_2}\varepsilon _N^{ - 1}} \right\rceil$, where ${\varepsilon _N}$ denotes the desired error of Newton iteration without considering the truncation error. We can also follow the result in \cite{CPP12} that the truncation error ${{\hat e}_{s'}}$ satisfies ${{\hat e}_{s'}}: = \left| {{{\hat z}^{\left( {s'} \right)}} - {z^{\left( {s'} \right)}}} \right| \le {s'}2^{-d}$.

       In short, with the number of the iteration steps being ${s'} = O\left( {{{\log }_2}d} \right)$, the error caused by the unitary ${\bf{U}}_{\sigma,\tau}$ is $\left| {z_j^{\left( {s'} \right)} - \frac{1}{{\sigma _j^2\beta _j^2}}} \right|  \le {\varepsilon _N} + {s'}2^{-d}$, where ${\varepsilon _N} \ge{2^{ - {2^{s'}}}}$.

\begin{figure}[ht]
\centering
\includegraphics[width=0.9\linewidth]{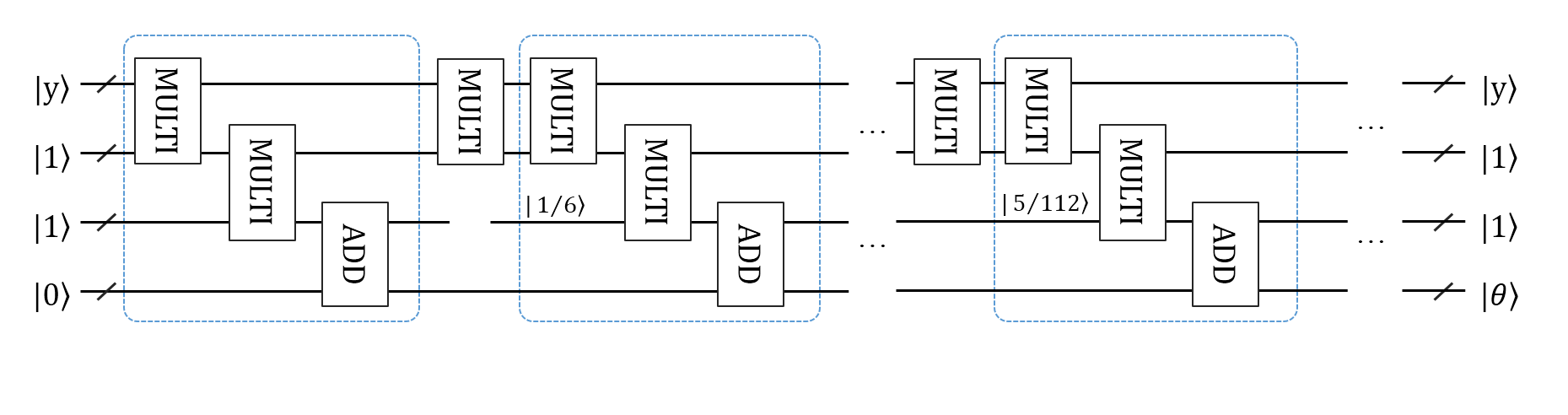}
\caption{The circuit for computing the rotation angle of ${c-R_y} $. }
\label{fig8}
\end{figure}

(2) For $c-R_y$, in order to make the output quantum state accurate, the rotation angle $\theta$ of $R_y$ satisfies $\theta= arc\sin\left ( y \right )$. 
Since $arcsin$ has a convergent Taylor series, we can approximate 

	\begin{equation}
	\begin{aligned}
		\theta = arcsin\left ( y \right ) \approx y+\frac{1}{6}y^{3} +\frac{3}{40}y^{5} +\frac{5}{112}y^{7} + \cdots .
	\end{aligned}
         \end{equation}	
Then the quantum circuit for computing the rotation angle of ${c-R_y}$ can be implemented as shown in Fig. \ref{fig8}. This circuit also only consists of the operations for addition and multiplication, therefore the number of qubits and gates required are $O(d)$ and $O\left[poly \left(d \right) \right]$ respectively.  

\begin{figure}[ht]
\centering
\includegraphics[width=0.5\linewidth]{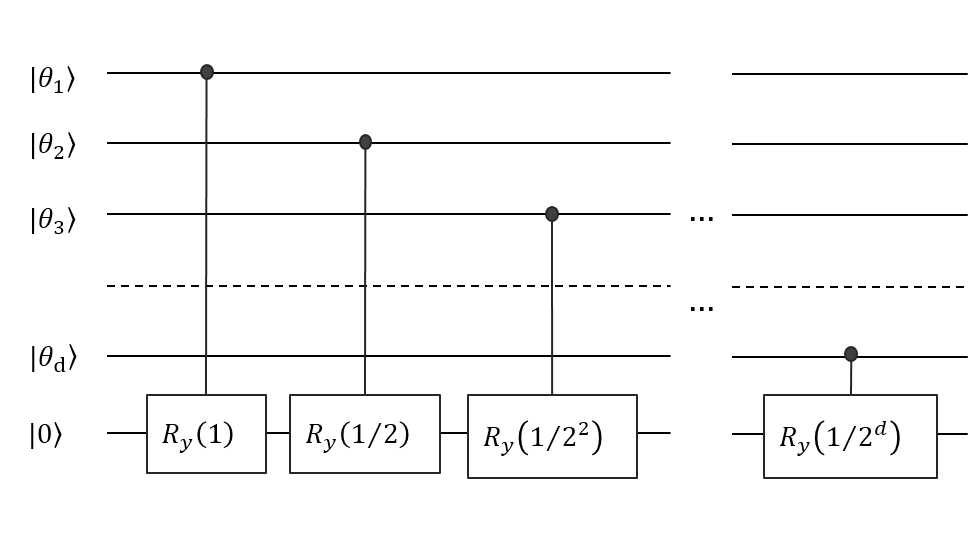}
\caption{The circuit for ${c-R_y} $. }
\label{fig9}
\end{figure}

The quantum circuit for $c-R_y$ is shown in Fig. \ref{fig9}, where $\theta_1, \cdots, \theta_d$ are the binary bits of the output $\theta$ in Fig. \ref{fig8}. Obviously, the space and time complexity are both $O(d)$.

\subsection {Complexity}
\label{sec4.4:level2}

We firstly analyze the space and time resources used in phase estimation. Let the efficient condition number of ${\widetilde {\bf{X}}} {\widetilde {\bf{X}}}^\dag $ be $\kappa$. As ${\textbf{A}^{(0)}}$ is induced by PCA of ${\widetilde {\bf{X}}}$, the condition number of ${\textbf{A}^{(0)}} {\textbf{A}^{(0)}}^\dag$ is not greater than $\kappa$, and so as the ${\textbf{A}^{(i)}} {\textbf{A}^{(i)}}^\dag$ for $i = 1, \cdots, s-1$. Therefore, ${\sigma _j}^2 \in \left[ {1/\kappa ,1} \right]$ and ${\beta _j}^2 \in \left[ {1/\kappa ,1} \right]$, and the numbers of qubits to store $\sigma _j^2$ in Reg. $B$ and $\beta _j^2$ in Reg. $C$ are both $b = O\left( {{{\log }_2}\kappa } \right)$. Moreover, we can learn from \cite{NC10} that $O\left( {{b^2}} \right)$ operations and two calls to the controlled-unitary black boxes are needed in the stage of phase estimation.

We now analyze the space and time complexity of the whole QAOP circuit. 
In one iteration $Q$ of the QAOP circuit, the number of qubits required for preparing the initial quantum state ${\left| {{\psi _{\textbf{A}^{(0)}}}} \right\rangle }$ is $O\left(p+q \right)=O\left [ \log_2 \left ( nk/ \epsilon \right )\right ]$. 
And phase estimation requires  $O\left( {{{\log }_2}\kappa } \right)$ qubits, where the condition number $\kappa$ is usually taken as $\kappa  = O\left( {1/\epsilon } \right)$. Therefore, the space complexity of phase estimation is $O\left( \log_2\epsilon^{-1} \right ) $. 
Taking the ancilla qubits in the controlled rotation into account, the number of qubits in this stage is $O(d+b)=O\left( {{{\log }_2}\kappa } \right)=O\left( \log_2\epsilon^{-1} \right )$. 
The number of qubits will not increase with the number of iterations of QAOP. To sum up, the total number of qubits required in the quantum circuit is $O\left [ \log_2 \left ( nk/ \epsilon \right )\right ]$.

Now turning to the time consumption. In one iteration $Q$, the number of gates for the initial state preparing stage is $O\left ( pq \right )=O\left [\log_2(nk)\log_2\epsilon^{-1} \right ] $. 
And phase estimation requires $ O\left( b^2 \right)= O\left[ \left( {{{\log }_2}\kappa } \right)^2 \right] = O\left[ \left( {{{\log }_2}\epsilon^{-1} } \right)^2 \right]$ operations and two calls to the controlled-unitary black boxes. The number of quantum gates in the controlled rotation is 
$O\left[s'poly \left(d \right) \right] = O\left[ s'{{poly}}\left({{\log }_2}\kappa \right)\right] = O\left[ s'{{poly}}\left({{\log }_2}\epsilon^{-1} \right)\right]$.
The number of iteration $Q$ is $s$, therefore the total time complexity of the QAOP circuit is $O[ ss' {\log_2(nk)} {{poly}}\left({{\log }_2}\epsilon^{-1} \right)]$.

In summary, $O\left[ {{{\log }_2}\left( {nk} /{\epsilon} \right)} \right]$ space and $O[  {\log_2(nk)}  {{poly}}\left({{\log }_2}\epsilon^{-1} \right)]$ elementary operations allow us to implement the QAOP circuit with fidelity at least $1-\epsilon$, when the number of iterations of the QAOP algorithm $s$ and the number of Newton's iteration $s'$ are both small constant numbers.

\section{Conclusions}
\label{sec5:level1}
We have shown that the proposed quantum algorithm QAOP can be used to speed up the learning process of an important dimensionality reduction algorithm in pattern recognition and machine learning. 
We firstly reformulated the original AOP algorithm, therefore the quantum-classical interactions during the quantum algorithm can be omitted. We then proposed the QAOP algorithm and investigated the quantum circuits for solving the QAOP algorithm. The detailed quantum circuits for preparing the input quantum state and the circuits for solving the controlled rotation are presented. The space and time complexity of the quantum circuit show that the number of the qubits and gates required are $O\left[ {{{\log }_2}\left( {nk} /{\epsilon} \right)} \right]$ and $O[{\log_2(nk)} {{poly}}\left({{\log }_2}\epsilon^{-1} \right)]$, respectively. The result shows that the QAOP algorithm and QAOP circuit for dimensionality reduction may motivate to conduct new investigations in quantum machine learning.

\begin{acknowledgments}
This work was supported by the Funding of National Natural Science Foundation of China (Grants No. 61571226 and No. 61701229), Natural Science Foundation of Jiangsu Province, China (Grant No. BK20170802), China Postdoctoral Science Foundation funded Project (Grants No. 2018M630557 and No. 2018T110499), Jiangsu Planned Projects for Postdoctoral Research Funds (Grant No. 1701139B), and the Nanjing University of Aeronautics and Astronautics PhD short-term visiting scholar project. The authors also acknowledge Michele Mosca for inspiring discussions.
\end{acknowledgments}

\appendix

\section {Proof of theorem 1}
\label{A1}
 \textbf{Proof:}  
 According to Eq. (\ref{eq:A0}), we have the initialization of ${{\bf{A}}^{\left( 0 \right)}} =\sum\limits_{j = 1}^k { {\beta _j^{\left( 0 \right)}} \left| {{{\bf{u}}_j}} \right\rangle \left\langle {\bf{j}} \right|} $, where ${\beta _j^{\left( 0 \right)}}=1$ for all $j$'s. Obviously, $\left\langle {{{{\bf{u}}_j}}} {\left | {\vphantom {{{{\bf{u}}_j}} {{{\bf{u}}_{j'}}}}}  \right. \kern-\nulldelimiterspace}  {{{{\bf{u}}_{j'}}}} \right\rangle  = 0$ when $j \ne j'$ and $\left\langle {{{{\bf{u}}_j}}} {\left | {\vphantom {{{{\bf{u}}_j}} {{{\bf{u}}_{j'}}}}}  \right. \kern-\nulldelimiterspace}  {{{{\bf{u}}_{j'}}}} \right\rangle  = 1$ when $j = j'$.
In the following, we simplify  ${{\bf{A}}^{\left( 0 \right)}}$ as ${\bf{A}}$.
Therefore, we have
               
                	\begin{equation}
			\begin{aligned}
                       {\widetilde {\bf{X}}^T}{\bf{A}}
                       & = \left( {\sum\limits_{j = 1}^r {{\sigma _j}\left| {{{\bf{v}}_j}} 
                       \right\rangle \left\langle {{{\bf{u}}_j}} \right|} } \right)
                       \left( {\sum\limits_{j' = 1}^k {\beta _{j'}^{\left( 0 \right)}} 
                       {\left| {{{\bf{u}}_{j'}}} \right\rangle 
                       \left\langle {\bf{j'}} \right|} } \right) \\
                       & = \sum\limits_{j = 1}^k {\sum\limits_{j' = 1}^k {{\sigma _j}{\beta _{j'}^{\left( 0 \right)}} 
                       \left| {{{\bf{v}}_j}} \right\rangle \left\langle {{{{\bf{u}}_j}}}
 			{\left | {\vphantom {{{{\bf{u}}_j}} {{{\bf{u}}_{j'}}}}}
			 \right. \kern-\nulldelimiterspace}
 			{{{{\bf{u}}_{j'}}}} \right\rangle \left\langle {\bf{j'}} \right|} }  
                       = \sum\limits_{j = 1}^k {{\sigma _j} 
                       {\beta _{j}^{\left( 0 \right)}} 
                       \left| {{{\bf{v}}_j}} \right\rangle \left\langle {\bf{j}} \right|} 
                       \end{aligned}
                    \label{eq:XA}
                	\end{equation}
and ${\widetilde {\bf{X}}^T}{\bf{A}}{{\bf{A}}^T}\widetilde {\bf{X}} = \sum\limits_{j = 1}^k { 
{\left( \sigma _j{\beta _j^{\left( 0 \right)}} \right)^2}
\left| {{{\bf{v}}_j}} \right\rangle \left\langle {{{\bf{v}}_j}} \right|} $. 
According to the Eq. (\ref{eq:B}), we have:
                	\begin{equation}
                      \begin{aligned}
				{\bf{B}} & 
                      		=  {\left( {{{\widetilde {\bf{X}}}^T}{\bf{A}}{{\bf{A}}^T}\widetilde {\bf{X}} 
                      		+ {\lambda _2}{\bf{I}}} \right)^{ - 1}}{\widetilde {\bf{X}}^T}{\bf{A}} \\
 				& = {\left( {\sum\limits_{j = 1}^k 
				{{\left( \sigma _j{\beta _j^{\left( 0 \right)}} \right)^2} \left| {{{\bf{v}}_j}} \right\rangle 
				\left\langle {{{\bf{v}}_j}} \right|}  + {\lambda _2}\sum\limits_{j = 1}^r 
				{\left| {{{\bf{v}}_j}} \right\rangle \left\langle {{{\bf{v}}_j}} \right|} } \right)^{ - 1}}
				\left( {\sum\limits_{j' = 1}^k {{\sigma _{j'}} {\beta _{j'}^{\left( 0 \right)}} 
				\left| {{{\bf{v}}_{j'}}} \right\rangle \left\langle {\bf{j'}} \right|} } \right)\\
 				& = {\left( {\sum\limits_{j = 1}^k {\left( 
				{{\left( \sigma _j{\beta _j^{\left( 0 \right)}} \right)^2}+ {\lambda _2}} \right)
				\left| {{{\bf{v}}_j}} \right\rangle \left\langle {{{\bf{v}}_j}} \right|} } \right)^{ - 1}}
				\left( {\sum\limits_{j' = 1}^k {{\sigma _{j'}} {\beta _{j'}^{\left( 0 \right)}} 
				\left| {{{\bf{v}}_{j'}}} \right\rangle 
				\left\langle {\bf{j'}} \right|} } \right)\\
 				& = \left( {\sum\limits_{j = 1}^k {{{\left( 
				{{\left( \sigma _j{\beta _j^{\left( 0 \right)}} \right)^2} + {\lambda _2}} \right)}^{ - 1}}
				\left| {{{\bf{v}}_j}} \right\rangle \left\langle {{{\bf{v}}_j}} \right|} } \right)
				\left( {\sum\limits_{j' = 1}^k {{\sigma _{j'}} {\beta _{j'}^{\left( 0 \right)}} 
				\left| {{{\bf{v}}_{j'}}} \right\rangle 
				\left\langle {\bf{j'}} \right|} } \right)\\
 				& = \sum\limits_{j = 1}^k {\sum\limits_{j'= 1}^k 
				{{{\left( {{\left( \sigma _j{\beta _j^{\left( 0 \right)}} \right)^2} + {\lambda _2}} \right)}^{ - 1}}
				{\sigma _{j'}} {\beta _{j'}^{\left( 0 \right)}} 
				\left| {{{\bf{v}}_j}} \right\rangle \left\langle {{{{\bf{v}}_j}}}
 				{\left | {\vphantom {{{{\bf{v}}_j}} {{{\bf{v}}_{j'}}}}}
				 \right. \kern-\nulldelimiterspace}
				 {{{{\bf{v}}_{j'}}}} \right\rangle \left\langle {\bf{j'}} \right|} }  \\
				& = \sum\limits_{j = 1}^k 
				{\frac{{{\sigma _j {\beta _{j}^{\left( 0 \right)}} }}}
				{{{\left( \sigma _j {\beta _j^{\left( 0 \right)}} \right)^2} + {\lambda _2}}}
				\left| {{{\bf{v}}_j}} \right\rangle \left\langle {\bf{j}} \right|} 
			\end{aligned}
                    \label{eq:B1}
                	\end{equation}

Similarly, we have 
                	\begin{equation}
                      \widetilde {\bf{X}}{\bf{B}} 
                      = \left( {\sum\limits_{j = 1}^k {{\sigma _j}\left| {{{\bf{u}}_j}} \right\rangle 
                      \left\langle {{{\bf{v}}_j}} \right|} } \right)
                      \left( \sum\limits_{j' = 1}^k 
				{\frac{{{\sigma _{j'} {\beta _{j'}^{\left( 0 \right)}} }}}
				{{{\left( \sigma _{j'} {\beta _{j'}^{\left( 0 \right)}} \right)^2} + {\lambda _2}}}
				\left| {{{\bf{v}}_{j'}}} \right\rangle \left\langle {\bf{j'}} \right|} \right) 
                      = \sum\limits_{j = 1}^k 
                      {\frac{{{\sigma _j^2 {\beta _{j}^{\left( 0 \right)}} }}}
				{{{\left( \sigma _j {\beta _j^{\left( 0 \right)}} \right)^2} + {\lambda _2}}}
                      \left| {{{\bf{u}}_j}} \right\rangle 
                      \left\langle {\bf{j}} \right|} 
                    \label{eq:XB}
                	\end{equation}
and 	$\widetilde {\bf{X}}{\bf{B}}{{\bf{B}}^T}{\widetilde {\bf{X}}^T} 
= \sum\limits_{j = 1}^k {
{\left (\frac{{{\sigma _j^2 {\beta _{j}^{\left( 0 \right)}} }}}
				{{{\left( \sigma _j {\beta _j^{\left( 0 \right)}} \right)^2} + {\lambda _2}}}  \right)^2}
\left| {{{\bf{u}}_j}} \right\rangle \left\langle {{{\bf{u}}_j}} \right|} $.

According to the Eq. (\ref{eq:A}), we have:

	        \begin{equation}
		\begin{aligned}
			{{\bf{A}}^{\left( 1 \right)}} 
			&= {\left( {\widetilde {\bf{X}}{\bf{B}}{{\bf{B}}^T}
			{{\widetilde {\bf{X}}}^T}} \right)^{ - 1}}\widetilde {\bf{X}}{\bf{B}}\\
			& = \left( {\sum\limits_{j = 1}^k {
			{\left (\frac {{{\left( \sigma _j {\beta _j^{\left( 0 \right)}} \right)^2} + {\lambda _2}}}
			{{{\sigma _j^2 {\beta _{j}^{\left( 0 \right)}} }}}
				  \right)^2}
			\left| {{{\bf{u}}_j}} \right\rangle \left\langle {{{\bf{u}}_j}} \right|} } \right)
			\left( {\sum\limits_{j' = 1}^k {
			\frac{{{\sigma _{j'}^2 {\beta _{j'}^{\left( 0 \right)}} }}}
				{{{\left( \sigma _{j'} {\beta _{j'}^{\left( 0 \right)}} \right)^2} + {\lambda _2}}}
			\left| {{{\bf{u}}_{j'}}} \right\rangle 
			\left\langle {\bf{j'}} \right|} } \right)\\
			& = \sum\limits_{j = 1}^k {\sum\limits_{j' = 1}^k {
			{\left (\frac {{{\left( \sigma _j {\beta _j^{\left( 0 \right)}} \right)^2} + {\lambda _2}}}
			{{{\sigma _j^2 {\beta _{j}^{\left( 0 \right)}} }}}
				  \right)^2}
			\frac{{{\sigma _{j'}^2 {\beta _{j'}^{\left( 0 \right)}} }}}
				{{{\left( \sigma _{j'} {\beta _{j'}^{\left( 0 \right)}} \right)^2} + {\lambda _2}}}
			\left| {{{\bf{u}}_j}} \right\rangle \left\langle {{{{\bf{u}}_j}}}
 			{\left | {\vphantom {{{{\bf{u}}_j}} {{{\bf{u}}_{j'}}}}}
			 \right. \kern-\nulldelimiterspace}
 			{{{{\bf{u}}_{j'}}}} \right\rangle \left\langle {\bf{j'}} \right|} } \\
			& = \sum\limits_{j = 1}^k { 
			\frac {{{\left( \sigma _j {\beta _j^{\left( 0 \right)}} \right)^2} + {\lambda _2}}}
			{{{\sigma _j^2 {\beta _{j}^{\left( 0 \right)}} }}} 
			\left| {{{\bf{u}}_j}} \right\rangle \left\langle {\bf{j}} \right|} 
		\end{aligned}
                 \label{eq:A1}
                	\end{equation}

Eq. (\ref{eq:A1}) shows that after one iteration of the algorithm, only the singular values of ${\bf{A}}$ are updated while the  singular vectors stay the same. 
Therefore, for $i$-th iteration of the algorithm, let  ${{\bf{A}}^{\left( i \right)}} =\sum\limits_{j = 1}^k { {\beta _j^{\left( i \right)}} \left| {{{\bf{u}}_j}} \right\rangle \left\langle {\bf{j}} \right|} $, where ${\beta _j^{\left( 0 \right)}}=1$ for all $j$'s according to Eq. (\ref{eq:A0}). We can easily have $\beta _j^{\left( i \right)} = \frac{{{{\left( {{\sigma _j}\beta _j^{\left( {i - 1} \right)}} \right)}^2} + {\lambda _2}}}{{\sigma _j^2\beta _j^{\left( {i - 1} \right)}}}$.

\qed 

\bibliography{QAOP}

\end{document}